\documentclass[]{spie}  %>>> use for US letter paper
\usepackage{amsmath,amsfonts,amssymb}
\usepackage{graphicx}
\usepackage[colorlinks=true, allcolors=blue]{hyperref}

\title{Real-time adaptive optics control with a high level programming language}

\author[a]{William Thompson}
\author[b]{Darryl Gamroth}
\author[a]{Christian Marois}
\author[a]{Olivier Lardière}
\affil[a]{National Research Council, 5071 W Saanich Rd, Victoria, Canada}
\affil[b]{Rubus Technologies Inc., Victoria, Canada}
\authorinfo{Further author information: (Send correspondence to W.T.)\\W.T..: E-mail: William.Thompson@nrc-cnrc.gc.ca}

\begin{document} 
\maketitle

\begin{abstract}
   Adaptive optics systems are usually prototyped in a convenient but slow language like MATLAB or Python, and then re-written from scratch using high-performance C/C++ to perform real-time control. This duplication of effort adds costs and slows the experimentation process. We present instead a technical demonstration of performing real time, sub-millisecond latency control with an adaptive optics system using the high-level Julia programming language. This open-source software demonstrates support for multiple cameras, pixel streaming, and network-transparency distributed computing. Furthermore, it is easy to interface it with other programming languages as desired.
\end{abstract}


\section{INTRODUCTION}
\label{sec:intro}  % \label{} allows reference to this section
Adaptive optics, or ``AO'', is an essential technique to modern astronomy. By compensating for the shimmering optical effects of the Earth's atmosphere, AO delivers the full resolution of large ground-based optical and near-infrared telescopes. Armed with AO, these telescopes have brought about a range of stunning scientific results, including the monitoring of stars orbiting the Milky Way's central super-massive black-hole and images of extrasolar planets orbiting nearby stars (a chief endeavor of these authors).

An adaptive optics system is powered by three main components. The first is a wavefront sensor (WFS), used to measure optical errors introduced by the Earth's atmosphere. The WFS is a combination of an optical device (e.g. a Shack-Hartmann or Pyramid) and a high-speed camera that is read at hundreds or thousands of frames per second. The second component of an AO system is one or more active optical elements---typically deformable mirrors (DMs)---which can, using a digital signal, apply arbitrary optical phase correction.  The third and final component  is the real time computer, or RTC. The RTC is the middle link that receives information from the wavefront sensor camera, performs some calculation, and then commands the deformable mirror to respond in an optimal way.

The RTC component must respond as quickly as possible to each camera frame. The main performance metrics aren't throughput (how much data can we process per minute), but latency and jitter (how quickly, and consistently quickly can we complete the calculations after receiving the data). 

The field of adaptive optics is now mature, and there are many high quality implementations operating on telescopes at any moment around the globe. Nonetheless, research into improved AO technology continues, as new telescopes, instruments, and scientific goals develop.
Many existing proprietary and open source RTC frameworks exist, including HEART \cite{heart}, CACAO \cite{cacao}, OOMAO\cite{oomao}, PyRTC\cite{pyrtc} and many others. 
This article will focus on the RTC component, and how leveraging new developments in programming language design and compilers can improve the velocity at which new adaptive optics systems and techniques are developed, tested, and deployed. This could ultimately help AO scientists and technicians deliver exciting new science capabilities faster than ever before.

\subsection{Current RTC Development Practices}

Before we continue, we make a brief note on terminology. In this article we will refer to AO scientists and to RTC engineers. We will imagine that these are two separate groups of people, with skills in either adaptive optics controls or in high performance programming. In practice, these lines are frequently blurred. Still, in our discussion we will treat them as separate for rhetorical convenience and because we imagine that these are somewhat different roles, even when conducted by the same individuals. 

We now consider the current state of affairs in real time controller (RTC) software development.
Typically when a new instrument or capability is proposed, the first step is to simulate the device. An AO scientist develops a model of the optical device, the wavefront sensor, and the deformable mirror. They then implement and consider different control schemes within this digital sandbox.
At this stage, the iteration speed of the simulation and control system are not a high priority. Instead, it is the iteration speed of the entire experimentation process that must be maximized. For this reason, the scientist typically implements their simulation and control experiments in a high level programming language like MATLAB, Python, or IDL. These programming languages have automatic memory management of some form, support interactive terminal use, offer convenient plotting functionality, and, at least in the case of MATLAB, provide convenient syntax for the linear algebra operations that feature centrally in AO control.

Later, they might advance to testing these ideas on a physical optical bench. This is often done using the same high level language code, combined with either a device control framework (e.g. \cite{}) or an ad-hoc system of scripts and device SDKs. At this stage, performance is often still not critically important. Although the system is tested with real optics and hardware, the scientist can often test their control software at reduced speed, perhaps on the order of tens or low-hundreds of Hz. Likewise, at this stage latency and jitter are often not critically important, and can be compensated by further reducing the loop speed or otherwise delaying control signals to stabilize loops speed to the millisecond level.

After their successful demonstration, AO scientist then
seeks to implement their new device or control algorithm on a real telescope. At this stage, they typically document the math and algorithms behind the control, and hand this specification off to one or more RTC engineers. The engineers are then responsible for the significant task of translating this specification (and perhaps some snippets of high-level code) into a compiled programming language that can reach the performance, latency, and jitter requirements necessary to keep up with the Earth's rapidly changing atmosphere. The compiled programming language chosen is always, to the best of our knowledge, either C or C++.

This final step in translating research code into a finished product is expensive. The RTC engineers must develop an understanding of the specification, must program and optimize the loop calculations, and also build up non-real time scaffolding around the loop code---that is, all of the work to load and prepare various control matrices, allocate buffers, connect and configure different hardware, etc.
This work proceeds in only one direction. Once the highly optimized C/C++ code is written, it may not be handed back to the AO scientist for verification, or for use in speeding up their future simulations. If either the AO scientist or RTC engineer determines that something about the algorithm should be improved, that change must be made to both simulation/lab and RTC software stacks, doubling the total amount of work done.

\subsection{Speaking the Same Language}

We wonder if now, thanks to new developments in programming language design and compilers, it might be possible to design a system where re-writing code in two languages is not necessary.
Such a programming language would have to offer both the convenient and high-level syntax and semantics that are necessary for rapid development, and also provide sufficient low-level control to reach real-time performance targets.

We expect that for peak performance, some components of an RTC framework will always require code that is carefully tuned for low-level control of hardware. We must therefore not only select a suitable programming language and runtime, but also design an RTC framework that isolates as much of this low-level complexity as possible.
In the best case, the AO scientist can be supported by the framework such that straightforward code performs well, and that such code can be passed off to an RTC specialist to make performance optimizations. The improved code could be returned to the AO scientist so that they can use it in future simulations or bench-top experiements, and now benefit from the improved performance. 

We note that others have had similar ideas. The \cite{} CACAO framework enables relatively straightforward data transmission between a real time component written in C/C++ and external software written e.g. in Python.  Most notably, the PyRTC framework\cite{pyrtc} implements a real time framework using Python and the Numba\cite{numba} JIT compiler.

\subsection{Julia}

The Julia programming language \cite{bezansonJuliaFastDynamic2012} was developed by the Julia Lab at MIT.
The purpose of Julia was to design a programming language with syntax and semantics that are both convenient for scientists, and amenable to compilation with a just-in-time (JIT) compiler. The choice of language semantics is important for high quality JIT compilation, as many existing programming languages include minor features that, while convenient, thwart attempts compile efficient computer code.

The Julia runtime uses the JIT module of the LLVM compiler toolkit---the same backend that powers the Clang C/C++ compiler and the main implementation of the Rust compiler. This choice of technology stack means that well-written Julia code can compile to essentially the same computer instructions as code written in C/C++. 

In addition to being able to write fast code in Julia, Julia code can interface very easily with existing libraries written in C or Python. 
Shared libraries compiled from C/C++  can be accessed  using the built-in \verb|ccall| functionality without having to compile a wrapper layer and with negligible overhead. On the opposite end of the spectrum, Julia can use Python libraries via the \verb|PythonCall.jl|\cite{pythoncall} or \verb|PyCall.jl|\cite{pycall} libraries. Using a Python library will of course cause Julia to operate as slowly as Python, but can be a very convenient way to access device SDKs already wrapped for Python in circumstances where ultra-high performance is unnecessary (e.g. operating a filter wheel or motorized stage).

The language semantics and syntax exposed by the Julia language are easy to learn and use. Programs written in MATLAB or Python/NumPy\cite{harris2020array} can often be translated into Julia quite easily, or written from scratch using its familiar math-inspired syntax.
Code written in what we would call ``high level Julia'' often perform somewhat faster than implementations in MATLAB or Python/NumPy, but usually exhibit very poor jitter.
Unlike MATLAB and Python/NumPy, however, in Julia it is very convenient to optimize this code in various ways to eliminate dynamic function calls and memory allocations. This kind of manual optimization process, best done by an experienced software developer, still results in Julia code that appears similar to the original ``high level'' implementation, but now performs with the same speed and jitter one would expect from a ground-up rewrite in C/C++.

There are a few important considerations about this optimization process for real-time applications that we now highlight.

\subsubsection{Type-Stability}

Julia is a dynamic programming language, i.e. a language in which the types of variables do not have to be specified in advance and can change during the execution of a program. This is similar to e.g. Python, where assigning \verb|x=1| at an interactive prompt (assigning the integer 1 to the label x), and then \verb|x=1.0| (assigning the floating point value 1.0 to the label x) is perfectly valid.

Type-stability is a concept in Julia that describes the ability of the compiler to statically determine the types of all variables used in a function for a given set of input types. Consider the following function:

\begin{verbatim}
function example_slow_1()
   # Generate a random number between 0 and 1
   x = rand()
   if x > 0.5
      result = 2    # A 64-bit signed integer
   else
      result = 3.5  # A 64-bit float
   end
   # Could be either an integer or a float randomly
   return result
end
\end{verbatim}

In the \verb|example_slow_1| function, the compiler will determine that the user wants either the integer 2 or the floating point value 3.5 at random. These two formats have different binary representations. The compiler may therefore box the value resulting in a dynamic memory allocation. This kind of dynamism is very useful when writing any kind of non-real time code---either when prototyping, or for the large portions of a codebase that are not time-sensitive.
That said, the dynamic allocations are a source of timing jitter in a real time loop and should be avoided. 
This can be accomplished by ensuring the code is \emph{type-stable}, as follows:
\begin{verbatim}
function example_fast_1()
   x = rand()
   if x > 0.5
      result = 2.0  # <-- now a float in both branches
   else
      result = 3.5
   end
   return result
end
\end{verbatim}

This subtle change allows the compiler to know the types of each variable ahead of time. With the standard Julia compiler/runtime, this results in devirtualization and allows the compiled function to use registers and stack-allocated memory, avoiding dynamic heap-allocations.

Notably, type-stability is a property of a given function (and the functions it calls) rather than an entire program. One can ensure that code called in a real time loop is type stable, and therefore has consistently high performance, without having to worry about setup code or error paths.

Various tooling exists to ensure that a given part of a Julia program is type stable. These include bundled code-introspection tools like \verb|code_warntype|, profilers that highlight dynamic dispatches (Julia's version of virtual function calls), and open-source libraries like \verb|DispatchDoctor.jl|\cite{dispatchdoctor},  \verb|JET.jl|\cite{jet}, and \verb|Cthulhu.jl|\cite{cthulhu}.

\subsubsection{Heap Allocations}
The standard Julia runtime uses a tracing garbage-collector (GC) to reclaim unused memory. When a memory allocation is required (for example to create a new array) and if the available heap memory has run low, the program is paused and the heap is scanned for allocated memory that is no longer accessible by any active variables. Any found memory is reclaimed, and the requested new allocation can then proceed.

Automatic memory management is helpful for high level code, but should be avoided in real time loops. If not, there will periodically be a pause while the GC scans and frees memory. This pause can be as short as a few hundred microseconds for small programs, or as long as several hundred milliseconds for large multi-threaded programs (as the GC must coordinate pausing all threads and scanning the entire shared heap).

Fortunately, heap allocations can be avoided by allocating all working arrays before starting a real time loop and ensuring that latency-sensitive code is type-stable. 
Consider the following two examples, the first with dynamic heap allocations occurring in each iteration, and the second with no heap allocations in the loop body:

\begin{verbatim}
function example_2_slow(camera_stream, interaction_matrix, dm_output)
   for camera_pixels in camera_stream
      # The `*' operator, if used with a matrix, performs a matrix multiplication using the
      # currently selected BLAS library
      new_dm_command = interaction_matrix * camera_pixels
      # The new_dm_command array is allocated each frame
      publish(dm_output, new_dm_command)
   end
end

function example_2_fast(camera_stream, interaction_matrix, dm_output)
   # We pre-allocate space for the new_dm_command once and reuse it on each iteration
   new_dm_command = zeros(size(interaction_matrix,1))
   for camera_pixels in camera_stream
      # In Julia, naming a function with an `!' at the end is a convention to indicate that
      # it operates in place, overwriting an argument:
      mul!(new_dm_command, interaction_matrix, camera_pixels)
      publish(dm_output, new_dm_command)
   end
end
\end{verbatim}

Both versions of this function are about as fast, but the second version using in-place operations does not allocate any memory during the loop and therefore never triggers any GC pauses.
Tooling exists to detect heap allocations in a given section of a Julia program, such as the included allocation profiler and the \verb|AllocCheck.jl|\cite{alloccheck} package.

\section{Software Architecture}

We now present the architecture for our Julia-based RTC framework. This framework supports the following features:
\begin{itemize}
   \item High performance, with low latency and jitter
   \item Robust and auditable implementations of pipeline stages described using hierarchical state machines
   \item Network transparency (computations can be freely spread across one or more servers without any software changes)
   \item Ability to swap various groups of loop parameters and matrices atomically between loop iterations
   \item Support for recording telemetry and control messages from any or all pipeline stages
   \item Ability to replay recorded telemetry or control messages
\end{itemize}

We developed this RTC framework for the SPIDERS instrument\cite{maroisDeploymentFocalPlane2022,2023aoel.confE..91L,10.1117/12.2629644,thompsonSPIDERS2024}, an upcoming high contrast imaging instrument that implements focal plane wavefront sensing, low order wavefront sensing, and a high-speed scanning imaging Fourier transform spectrograph\cite{iftsAdam}. 

The software we describe is published as a mix of MIT-licensed building blocks, available-by-request driver wrappers (sharing these publicly would in some cases breach supplier license agreements), and a few components that are currently proprietary. 
The software is available at the following URL: \url{https://github.com/New-Earth-Lab}.

% A few special considerations for the Julia language and runtime where taken in architecting this framework.
% % We present an RTC software stack that is designed to leverage the strengths of Julia, while guiding users away from coding patterns that could hurt overall RTC performance and latency.
% Based on experience with a previous Julia RTC software architecture, we have settled on a design that
% \begin{itemize}
%    \item separates setup and input/output (IO) operations from the critical real time loops,
%    \item guides users away from allocating temporary arrays on each loop iteration,
%    \item and mitigates garbage collector pauses from any accidental memory allocations in said loops.
% \end{itemize}

% % We split the RTC into a series of microservices, each responsible for one step of the RTC pipeline. 

\subsection{Control Loop Microservices}

We split our RTC framework into a set of single-threaded services that each carry out one stage of the RTC pipeline. Examples of these services might include receiving data from a camera, calibrating pixels, or performing matrix-vector multiplications. We separate these loops into isolated OS processes both for developer convenience and as a performance mitigation for unoptimized code.

The convenience aspect is that one can freely inspect or record the inputs/outputs of each pipeline stage, and start, stop, modify, and restart each step of the pipeline independently. As we will discuss later, this approach also allows an RTC pipeline to be easily split across multiple servers without any software changes.

The performance mitigation aspect of this architecture comes from Julia's current use of a stop-the-world garbage collector (GC).  Automatic memory management, like a GC, is a key feature of high level languages. A GC prevents unbounded memory growth without burdening programmers with manual memory management. This is accomplished by scanning the process for unused memory which can be safely reclaimed.
In Julia's case, the GC must momentarily pause all threads in a given process in order to perform this scan. This  means that if multiple loops are running on separate threads within the same process, all threads must coordinate to pause at the same time, after which the process's entire memory heap can be scanned. For a large multi-threaded Julia program that hasn't been optimized, this could take as much as 200-300ms, and can occurr every few minutes. 
This represents an unacceptable pause in a real time loop.
By contrast, the GC can operate on Julia processes with a single thread (and correspondingly smaller memory heap) in as little as 400 microseconds. 
Of course, well-written and optimized RTC code should have no dynamic memory allocations in the real time loop, removing the need for any GC pauses. Still, it is easy for such allocations to sneak in during development. This isolated process design mitigates the impact of any memory allocations until they can be detected and patched\footnote{Various software tools allow one to detect such allocations, either at compile time (\texttt{AllocCheck.jl}) or run time (the built-in profiling tools).}.

\subsection{Event-Driven Design}

\begin{figure*}
   \centering
   \includegraphics[width=\columnwidth]{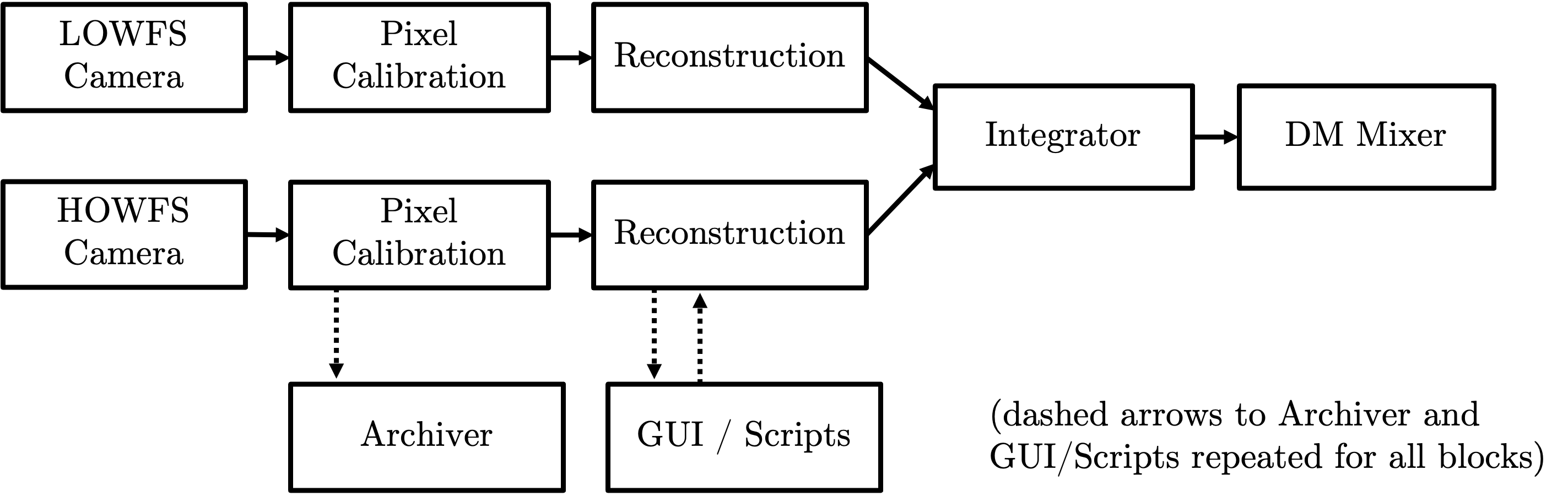}
   \vspace{8pt}
   \caption{
      Schematic of the RTC pipeline configured for the SPIDERS instrument\cite{maroisDeploymentFocalPlane2022,2023aoel.confE..91L,10.1117/12.2629644,thompsonSPIDERS2024}.
   }\label{fig:gui-cropped}
\end{figure*}

We use the open-source Aeron protocol and C implementation to communicate with and between services. The Aeron protocol originates from the high-frequency trading sector and development is currently funded by Adaptive Financial Consulting. It provides a publish/subscribe mechanism on any number of ``streams'' that operate over the network using UDP or between processes on the same computer using shared memory.
Aeron consists of a shared library used within each communicating process and a media driver, usually running externally, that coordinates the movement of data between processes or computers. 
We wrap the \verb|aeron| C library with our open-source \verb|Aeron.jl|\cite{aeron} wrapper so that it can be used conveniently from Julia.

Aeron is designed for very low latency and small messages can be transmitted in a single UDP packet. Aeron is designed to be entirely lock-free, up until the point of making a system call to transmit data over the network. For local communication between processes, shared memory is used in place of UDP packets and locks are avoided completely.
Aeron can support messages up to ~16Mb in size, broken up into multiple fragments if needed. For messages consisting of only a single fragment, Aeron can operate with no memory copying\footnote{The publisher requests a range of memory from aeron, fills this as needed, and then aeron coordinates making this same range of memory available to the subscribers}.

We adopt Simple Binary Encoding (SBE)\cite{sbe} as the serialization format used in our RTC system. This minimal format allows for the encoding of structures and arrays in a binary, cross-platform, and language agnostic way. The SBE format can allow zero-copy access to data structures. That is, in most implementations no memory needs to be copied to perform serialization/deserialization, and can be accessed in place. 

For our RTC framework, we specify two key types of message in SBE format. The first is a \verb|TensorMessage|, which encodes a block of one- to four-dimensional data with metadata for its size, offset, and so-forth. These \verb|TensorMessage|s are used to transmit camera frames and DM commands.
The second message format is \verb|EventMessage|, which indicates an event with a given (string) name has occurred. \verb|EventMessage|s carry an optional payload that can be a number, string, or another nested message. We use this mechanism to communicate new parameter values to services. In the case of e.g. interaction matrices, the matrix is encoded as a \verb|TensorMessage| and placed as the payload of an \verb|EventMessage|.
All messages share a common header with metadata including a description, the time a message was generated in a system, and correlation number. The time and correlation number are propagated through each step of the pipeline to track latency and the flow of data through the system.

Each service subscribes to an event stream and publishes a status stream. Received and processed event messages are republished on the status stream for logging purposes. Services may additionally subscribe to and/or publish one or more incoming or outgoing data streams of \verb|TensorMessage|s.

Within a single \verb|aeron| message (consisting of one or more fragments), we adopt the convention that multiple \verb|EventMessage|s can be concatenated together. If multiple events are received in a single \verb|aeron| message, they are processed in order before the next data message is processed. This allows for changing multiple service parameters atomically between loop iterations.

\subsection{Pixel Streaming}

Pixel streaming (processing data line by line, as it is received by the computer) can be implemented using the above \verb|TensorMessage| format with essentially no modifications. Tensor messages include metadata that specifies the location and offset of the data payload in up to four axes. The publisher can use these fields to then publish data one chunk at a time, specifying the location and offset of that chunk. 
With this format, a publisher can share a partial row of a camera frame, a complete row, or multiple complete rows, but not a mix of complete rows and partial rows.

As an additional optimization, the publisher can use the \verb|tryClaim| functionality from Aeron to request direct access to range of shared memory. The publisher can then populate this memory range and it will be shared with the subscriber with zero memory copies. This can be more friendly to the CPU cache because only a partial frame needs to be handled at a time.

Although all the mechanisms to support pixel streaming are in place, we do not currently use it in our RTC for the SPIDERS instrument as the required latency is easily met when transmitting full frames.

\subsection{Hierarchical State Machines}

We provide a library \verb|Hsm.jl|\cite{hsm} that implements hierarchical state machines in Julia. A hierarchical state machine is a way of representing a finite set of program states, including nested states.
We find that hierarchical state machines are a useful formalism for robustly implenting the services used in our framework, but note that it isn't a necessity---it is possible to get started writing a new RTC service with nothing more than a Julia \verb|for| loop receiving camera frames and publishing calculations.

To implement a service as a hierarchical state machine, one implements a specification of:
\begin{itemize}
   \item the set of possible program states
   \item the code to execute in a given state whenever an event is received
   \item the code to execute when preparing to enter a new state
   \item the code to execute when leaving a state
\end{itemize}

When an event is received by a service, it is dispatched to the state machine. Events can either by handled (e.g. when receiving incoming \verb|TensorMessage|s in a ``Processing'' state), or ignored and allowed to bubble up to a parent state (e.g. a ``Reset'' event can be handled in an outer state). Some events may trigger a state transition. The \verb|Hsm.jl| library then traverses the graph of possible states and tears down, or sets up variables as needed.

\begin{figure}
   \centering
   \includegraphics[width=250pt]{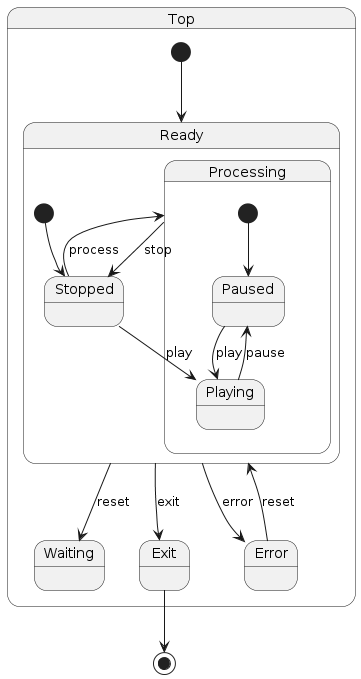}
   \caption{
      A visual representation of a hierarchical state machine. We re-use the pictured state machine for
      all the blocks of the main AO correction pipeline.
   }\label{fig:hsm}
\end{figure}

We standardized on a single state machine for each stage of the AO correction pipeline, shown in Figure \ref{fig:hsm}. Other services like device drivers implement their own state machines as is convenient. When the service is first started, it enters the ``Waiting'' state until it receives any needed parameters such as calibration images or reconstructor matrices. Once these are provided, it transitions into the ``Ready'' state, and then into the ``Processing'' state. In the ``Processing'' state, the service is subscribed to any incoming data streams and performs calculations. If in the ``Playing'' child state, the resulting calculations are published to the next stage of the pipeline. The ``Paused'' state is useful because it allows the user to momentarily stop sending updates without otherwise stopping or resetting the service. The Paused state still processes data so that the CPU cache is alreay warmed up, and we can therefore begin correcting as soon as the ``Playing'' state is entered without any delay. The ``Stopped'' state on the other hand ceases processing of incoming data and clears and existing state (for example, an integrated command).

% \subsection{Command Line Interface}
% We developed a command line tool that can send and listen for messages on a given aeron stream. This is useful for interacting with the system from a terminal interface, and for some scripting tasks. 

\subsection{Graphical User Interface}

\begin{figure*}
   \centering
   \includegraphics[width=\columnwidth]{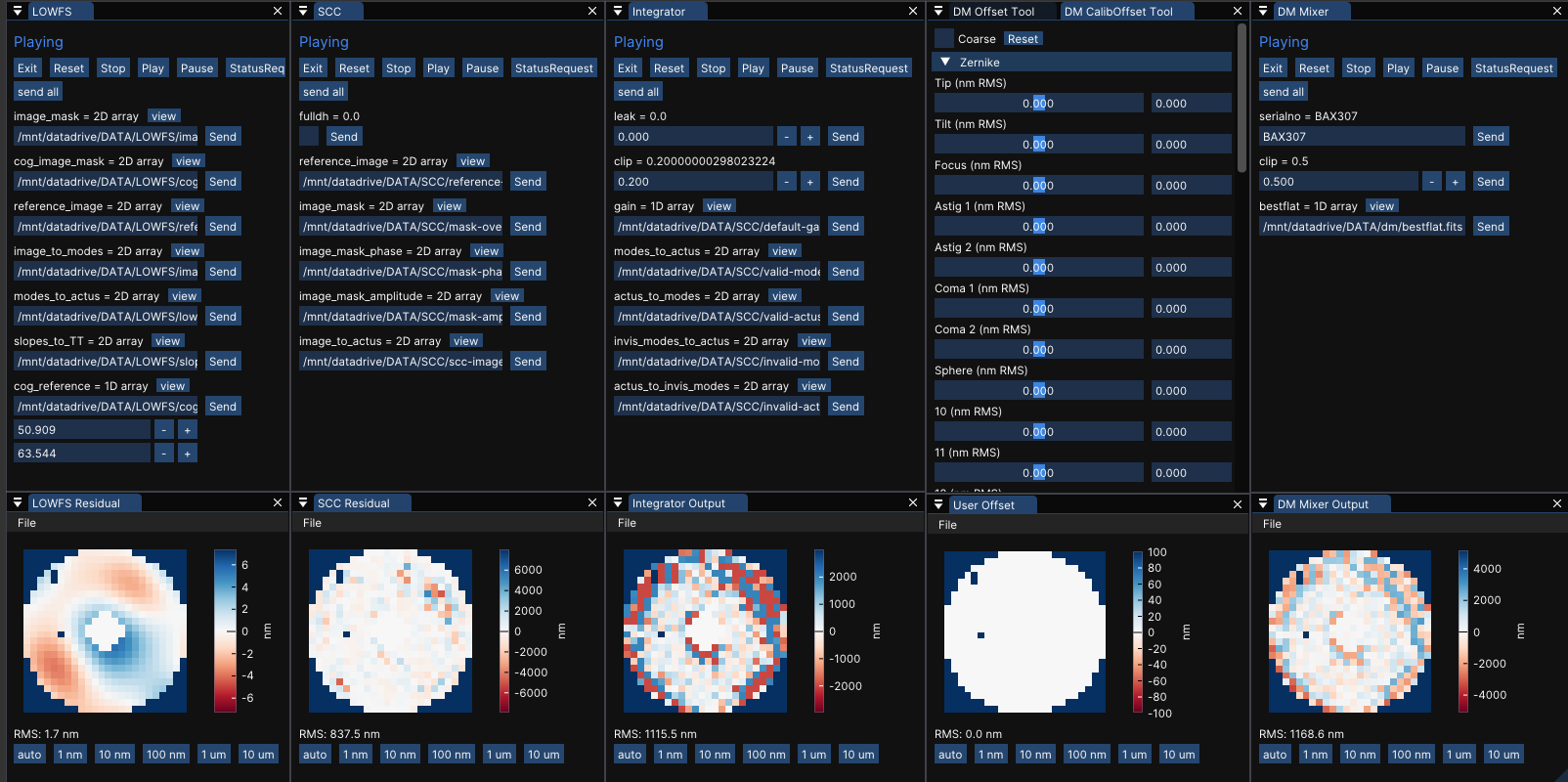}
   \caption{
      A zoomed in portion of Figure \ref{fig:gui-cropped} showing panels for some of the RTC pipeline stages.
   }\label{fig:pipeline-gui}
\end{figure*}

We developed a graphical user interface (GUI) to let the user interact and monitor the system. We developed the GUI using the immediate-mode GUI library, Dear ImGui\cite{imgui}. Dear ImGui is very convenient for making custom interfaces that render at 60 frame/s and 4k resolution using a GPU. As a downside for this convenience, Dear ImGui does not resemble a native OS application nor does it provide affordances for accessibility. We feel that this is acceptable since the RTC software can be controlled via a text-based command line interface, for scripted or interactive use.
Screenshots of the GUI are presented in Figures \ref{fig:gui-cropped} and \ref{fig:pipeline-gui}.

The GUI communicates with the RTC services using \verb|aeron|. When users interact with various controls, the GUI generates \verb|EventMessage|s that are sent to the appropriate service. When a service's state changes for any reason---either in response to a botton press, outside control, or an unexpected error---the service generates \verb|EventMessage|s that are received by the GUI and used to update the current display.
Likewise, the GUI subscribes to various data streams of \verb|TensorMessage|s and displays the most recent array using a heatmap or other plot.

\begin{figure*}
   \centering
   \includegraphics[width=\columnwidth]{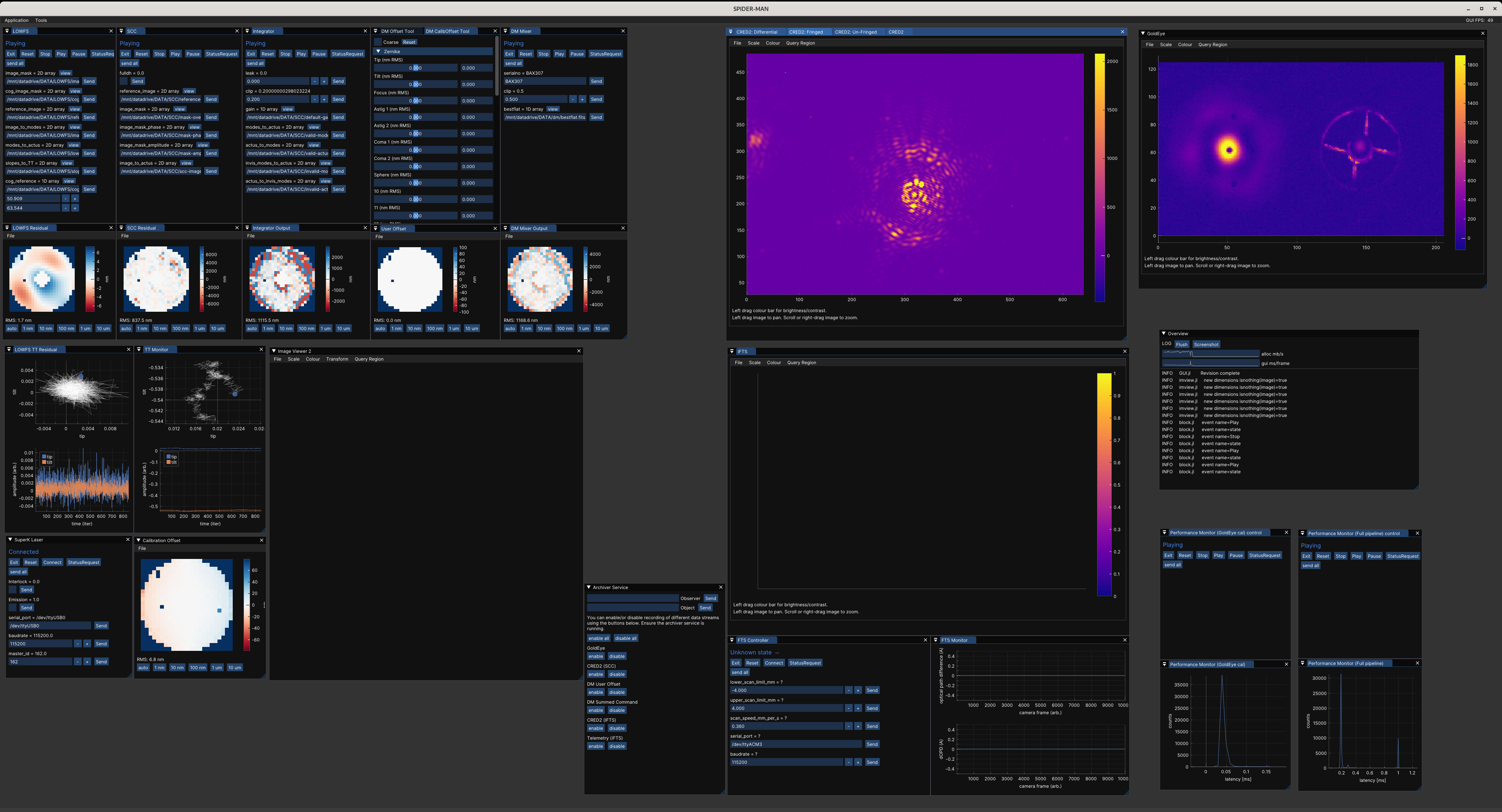}
   \caption{
      Screenshot of the Julia-based RTC GUI configured for the SPIDERS instrument(see Section \ref{sec:SPIDERS}). The top left panels show controls and DM output for different stages of the RTC pipeline. The bottom middle region shows the residual and integrated tip/tilt commands. The right top region shows two camera streams used for wavefront sensing. The bottom center shows a panel to control the Archiver, and panels for an Imaging Fourier Transform Spectrograph, one of the science instruments. The bottom right area shows a real time latency histgorams (see the results section for properly measured data).
   }\label{fig:gui-cropped}
\end{figure*}

\subsection{Telemetry Storage and Replay}

We implemented a service we call the Archiver that can subscribe to any of \verb|aeron| streams and record their contents to disk. We currently use a hybrid format of an SQLite database created for each night of observations and a set of flat binary files in SBE format.
When each message is received by the archiver, the full message is immediately appended to a flat binary file as-is. A message ``receipt'' is then stored in the SQLite database which includes the name of the binary file, the start and end bytes of the message, and the deserialized header of the message (recall that we use a header with the same metadata for all message formats used in the system).

This archive can then be examined to search for messages from a given stream number, and filtered via any metadata from the message headers. The resulting messages can be loaded from disk via memory mapping and either dumped into a more familiar format like FITS, or republished with a new time-stamp and correlation number.

Since we use the same mechanism to transport data, system events, and service parameters, one archive can contain all information needed to replay the complete state of the system for debugging, performance characterization, and more. We use further use this system for the transmission and archiving of science data.
The SQLite index can be used to time-align streams from different wavefront sensors and connect these with any given science frame, and e.g. the gain reconstructor matrices used by the AO loops.

For calibration and other non-real time tasks, we access camera streams via the archive. The script generates a series of DM commands, controls light sources, etc. while recording the current time-stamp at each step. At the end of the script, data matching those time stamps is queried from the archive.
This setup obviates the need to keep up with the live, high-speed data stream.

\section{Results}\label{sec:SPIDERS}

We now present timing results of our Julia-based RTC framework. We implement the complete control system for SPIDERS (the Subaru Pathfinder Instrument for Detecting Exoplanets and Retrieving Spectra; \cite{maroisDeploymentFocalPlane2022,thompsonPerformanceFastAtmospheric2022})
SPIDERS includes two wavefront sensors, two science cameras, a single deformable mirror, and a myriad of mechanisms, light sources, etc. We implement control for all of these devices in Julia, with the exception of the cameras. These are currently implemented using the Go language, though we intend to migrate them to Julia in the near future.
We benchmark the low order wavefront sensor loop only, as it operates at a much higher framerate than the high order wavefront sensor loop.

The low order wavefront sensor loop has the following key properties:
\begin{itemize}
   \item $206 \times 125$ pixels
   \item 1000 frame/s
   \item 468 actuators
\end{itemize}
The camera used is a Goldeye G-008 Cool, and the deformable mirror is an ALPAO mirror with 468 actuators arranged in a circular pupil.

We benchmarked the RTC performance on a server with 
a 32-core AMD Ryzen Threadripper PRO 3975WX.
The operating system was Rocky Linux version 9 with kernel version 5.14.0 and the \verb|PREEMPT_RT| real-time linux patch. The different RTC services were pinned to cores, but no real-time priority was set on their threads.  No further consideration was given to preventing other non-real time processes from being scheduled on the pinned cores, nor to limiting other background activities on the server. The GUI was active and running in a GNOME based environment.

\begin{figure*}
   \centering
   \includegraphics[width=0.49\columnwidth]{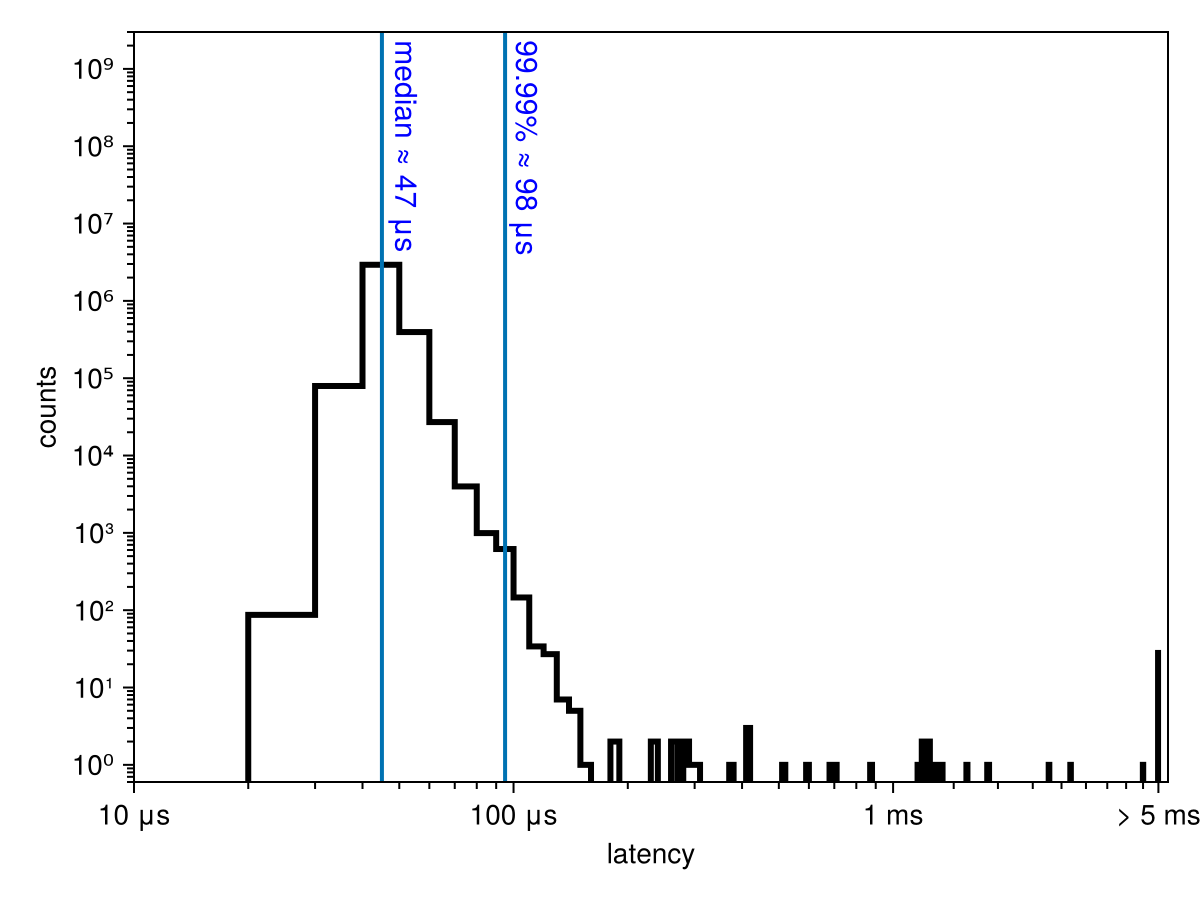}
   \includegraphics[width=0.49\columnwidth]{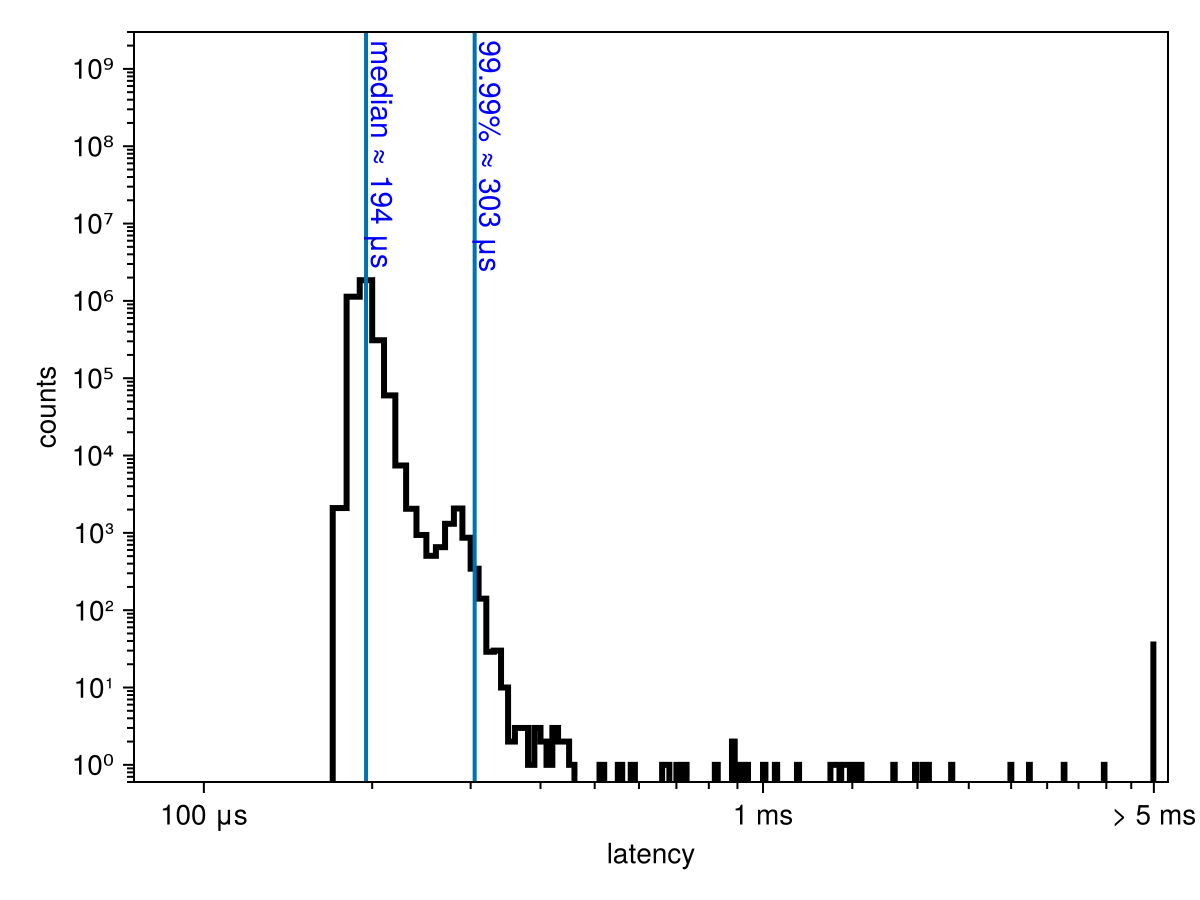}
   \caption{
      Performance characterization of the SPDIERS low order wavefront sensor. \textbf{Left:} latency measured from time the frame is finished being received by the RTC, through pixel calibration, and then received by the performance monitoring service. \textbf{Right:} latency measured from time the frame is finished being received by the RTC, through pixel calibration, wavefront reconstruction, command received by the integrator, integrated command received by the DM service, hardware command sent, and finally republished and received by the performance monitoring service.
   }\label{fig:latency}
\end{figure*}

The results of our performance monitoring are presented in Figure \ref{fig:latency}. 
We measure both the latency for frame calibration, and the latency for the complete pipeline from image arrival to DM command sent.
We measure latency with the start time recorded just after the entire frame has arrived in memory, and top time after information is received by a performance monitoring service. This extra transmission step to arrive in the performance monitoring service (after being sent to the DM) adds some additional latency, so this can be considered an upper bound. 
In the presented histograms, we consider any frame that arrives later than 1ms to be late.

We find that the Julia-based RTC framework can successfully keep up with real time tasks. 
After recording the latency of the full correction pipeline for a period of 56 minutes, only a fraction of 0.000017 frames complete processing after more than 1 ms.

We monitoried the Julia garbage collector in all tasks, and verified that it was not triggered a single time in the steady stage. That is, not a single GC pause occured during the recording of these measurements. We therefore conclude that the small tail of late frames are most likely a result of OS scheduling and/or hardware interrupts. These could be partially mitigated by more careful configuration of the operating system and hardware, and these considerations would apply to any choice of programming language.

\section{Conclusion}

In conclusion, we have demonstrated that Julia is a practical choice for implementing real time control for adaptive optics. We find that it can essentially reach the performance of C, and that jitter from the Julia runtime can be entirely avoided in real time loops. In addition, we find that the code is compact, easy to read and maintain, and could foster improved collaboration between AO scientists and RTC engineers. 

\acknowledgments % equivalent to \section*{ACKNOWLEDGMENTS}       

% References
\bibliography{report} % bibliography data in report.bib
\bibliographystyle{spiebib} % makes bibtex use spiebib.bst

\end{document}